\documentclass[fleqn,10pt]{wlscirep}

\usepackage{subfigure}
\usepackage{float}
\usepackage{lipsum}

\title{Refractory period in network models of excitable nodes: self-sustaining stable dynamics, extended scaling region and oscillatory behavior}

\author[1,2,+]{S. Amin Moosavi}
\author[1,*,+]{Afshin Montakhab}
\author[2]{Alireza Valizadeh}
\affil[1]{Department of Physics, College of Sciences, Shiraz University, Shiraz 71946-84795, Iran}
\affil[2]{Department of Physics, Institute for Advanced Studies in Basic sciences (IASBS), Zanjan 45137-66731, Iran}

\affil[*]{montakhab@shirazu.ac.ir}

\affil[+]{these authors contributed equally to this work}

\begin{abstract}
Networks of excitable nodes have recently attracted much attention
particularly in regards to neuronal dynamics, where criticality
has been argued to be a fundamental property. Refractory behavior,
which limits the excitability of neurons is thought to be an
important dynamical property. We therefore consider a simple model
of excitable nodes which is known to exhibit a transition to
instability at a critical point ($\lambda=1$), and introduce
refractory period into its dynamics.  We use mean-field analytical
calculations as well as numerical simulations to calculate the
activity dependent branching ratio that is useful to characterize
the behavior of critical systems. We also define avalanches and
calculate probability distribution of their size and duration. We
find that in the presence of refractory period the dynamics
stabilizes while various parameter regimes become accessible. A
sub-critical regime with $\lambda<1.0$, a standard critical
behavior with exponents close to critical branching process for
$\lambda=1$, a regime with $1<\lambda<2$ that exhibits an
interesting scaling behavior, and an oscillating regime with
$\lambda>2.0$. We have therefore shown that refractory behavior
leads to a wide range of scaling as well as periodic behavior
which are relevant to real neuronal dynamics.
\end{abstract}
\begin{document}

\flushbottom
\maketitle
\thispagestyle{empty}

%\noindent Please note: Abbreviations should be introduced at the first mention in the main text – no abbreviations lists. Suggested structure of main text (not enforced) is provided below.

\section*{Introduction}

Various physical, biological and chemical systems are composed of
interacting excitable agents and thus networks of excitable nodes
are widely used to model the behavior of such systems. Examples of
such systems include tectonic plates \cite{BuK,OFC}, Neural
networks \cite{BP1,SYYRP,SYPRP,HB,Garcia}, models of self-organized
criticality (SOC) \cite{BTW1,BTW2,B,P}, and epidemic (contagion)
spreading \cite{Grassberger,Karrer,Mieghem,Montakhab,Manshour}.
Oftentimes excitable nodes are modeled with threshold dynamics as
in the case of sandpile models of SOC which are thought to
underlie the wide range of scale-invariant behavior seen in
Nature.

Criticality in cortical dynamics is by now a widely studied and
well established field \cite{Plenz}. Neuronal avalanches have been
reported as collective scale-invariant behavior of neurons in the
cortical layers of mammalian brain
\cite{BP1,Plenz,PT,BP2,FIBSLD,PTLNCP,TBFC,SACHHSCBP,HTBC,C}.
Probability distribution functions of duration and size of the
neuronal avalanches are power law functions observed in a wide
range of space and time which are thought to be hallmarks of
critical systems. In addition to avalanche statistics, branching
ratio has also been used to characterize various time-series in
order to establish critical dynamics of the brain \cite{BP1}.
Criticality of the brain is therefore the subject of a myriad of
theoretical as well as experimental studies
\cite{BN,DanteC,BR,LHG,MMKN,APH,Beggs,AMIN,AMIN1}. Critical
dynamics is believed to underlie many functional advantages in a
healthy brain, including learning \cite{AH}, optimal dynamic range
\cite{SYPRP,KC,LSR1,LSR2,PTYJZZ}, efficient information processing
\cite{Beggs}, as well as optimal transmission and storage of
information \cite{SYYRP}.

Many excitable agents often display refractory behavior. This
behavior which is characterized by a time scale (i.e. refractory
period) is a time during which the excited agent cannot be
re-excited. The presence of such refractory period can clearly
affect the collective dynamics of excitable nodes. Neuronal
systems are perfect examples of networks of excitable nodes with
refractory period \cite{Kandel}.

A particularly useful model of excitable agents considers a random
network of such nodes with quenched excitatory connection weights
and probabilistic dynamics. It is well-known that such a model
exhibits a phase transition between stable and unstable regimes as
the average weight of connections is increased
\cite{LSR1,LSR2,LCOR}. More recently, it has been shown that the
addition of inhibitory connections leads to a ceaseless dynamics
which exhibits critical behavior by fine-tuning the system to the
transition point associated with the mentioned instability
\cite{LSOSR}. This fine-tuning essentially leads to an intricate
balance between excitatory and inhibitory connections which may
not be \textit{a priori} available and thus a non-generic
behavior. Here, we propose to study the original model (without
inhibitory connections) in the presence of refractory period.
Interestingly, we find that refractory period leads to stable
dynamics in the entire range of parameter. However, and perhaps
more interestingly, we are able to identify a critical point with
robust finite-size scaling behavior, a wide critical-like regime
with interesting scaling behavior, as well as a regime with
periodic behavior. Therefore, we show that refractory period in
addition to stabilizing the dynamics leads to a wide range of
parameters which show scaling or periodic behavior which are
hallmarks of real neuronal dynamics.

This article is organized as follows: The following section
discusses the model. Next, we show our analytical and numerical
results, respectively. We close by providing concluding remarks.

\section*{Model}

The model consists of $N$ excitable nodes on a random directed
graph where every two nodes are connected with probability $q$.
The average out-degree \textit{and} in-degree of the network is
equal to $\langle k \rangle=qN$. Weights of the connections
($w_{ij}$), that form the adjacency matrix of the network, are
randomly chosen real numbers in the range of $[0,2\sigma]$ with
the average connection weight of $\sigma$. If node $i$ is not
connected to node $j$, their connection weight is set to zero
($w_{ij}=0$). Every node can be in one of active or quiescent
states. Activity of the network is shown by the spatio-temporal
binary variable, $A_{i}(t)$, i.e. if the node $i$ is active at
time $t$ then $A_{i}(t)=1$ and when the node is quiescent
$A_{i}(t)=0$. The probability of a node to be activated at time
$t+1$ is equal to
\begin{equation}
\label{Eq1}
p\big(A_{i}(t+1)=1\big)=\delta_{0,A_{i}(t)} f\big(\sum_{j=1}^{N} w_{ij}A_{j}(t)\big)
\end{equation}
where $\delta_{0,A_{i}(t)}$ is the Kronecker delta, which is equal
to zero (one) if $A_{i}(t)=1$ $(0)$, implying refractory period of
one time step, i.e. the activation probability of a node that is
active at time $t$ will be equal to zero at $t+1$. $f$ is a
transfer function that transfers the total input of a node
($\sum_{j=1}^{N} w_{ij}A_{j}(t)$) at time $t$ to the activation
probability of that node at time $t+1$, and is chosen to be
\begin{equation}
\label{Eq2}
f(y)=
\begin{cases}
   y, & 0\leq y\leq 1\\
   1, & y>1.
\end{cases}
\end{equation}

In this work, our focus is on the aggregate activity of the system
which is defined as the number of active nodes at each time step
and is equal to $x_{t}=\sum_{i=1}^{N}A_{i}(t)$. We use $x_{t}$ as
a dynamical variable to analyze stability as well as statistical
properties of the system.

Before presenting our results, it is instructive to remark on
properties of the model without a refractory period. It is well
known that the behavior of this model, without a refractory
period, is governed by the largest eigenvalue of the adjacency
matrix \cite{LSR1,LSR2,LCOR,PTYJZZ} which is equal to
$\lambda=\sigma \langle k \rangle$ for the random graph that we
use \cite{LCOR,ROH}. It has been shown that this system exhibits
stability and instability in activity for $\lambda\leq1$ and
$\lambda>1$, respectively. In the case of stability, the system
requires external drive to be activated and $x=0$ is the stable
attractor of the dynamics. When unstable ($\lambda>1$) the
activity of the system increases and saturates at $x=N$. The
critical point of the system is $\lambda=1$ where the system
undergoes a transition from stability to instability. Poised at
the critical point, the system exhibits scaling behavior for
statistical properties of cascades of activity (avalanches) that
start by an external perturbation. Our main goal in this paper is
to scrutinize the behavior of the system when refractory period is
introduced into the dynamics, i.e. delta function in Eq.
(\ref{Eq1}). We are particularly interested in stability of
dynamics and/or whether generic scaling behavior similar to
cortical samples could arise in the model.

\section*{Results}

\subsection*{Mean-field analysis}
Dynamics of the system with refractory period is governed by Eq.
(\ref{Eq1}). It is clear that the dynamics is strongly affected by
the interaction weights ($w_{ij}$). We can roughly explain the
behavior of the system by considering different extremes of
$\lambda$ which is proportional to the average weight of
connections $\sigma$. For small values of $\lambda\ll 1$ the
interaction weights are small and any activity that starts by an
initial perturbation is expected to die out very fast leading to a
stability of the fixed point $x=0$. In the other extreme, for
large values of $\lambda \gg 1$, due to large values of
interaction weights the probability of firing is expected to be
equal to one for any node that receives input and is also
quiescent at the time (see Eq. (\ref{Eq1})). Therefore, at any
given time, after a transient period, there are two sets of nodes:
$x$ that are active, and $N-x$ that are in refractory period and
will be active in next time step. Clearly, the system exhibits
oscillations in this limit where all active nodes become quiescent
and vice versa. We generally expect that there exists an important
range of $\lambda$ over which the system changes its behavior from
a ceasing stable phase to a cease-less periodic phase. It is
possible that the transition passes though a critical point or
region.

In order to analyze the behavior of the model we use the aggregate
activity of the system $x_{t}$ as a function of time, and
calculate the \textit{activity dependent} branching ratio $b(M)$
\cite{MSP} which is the expectation value of $x_{t+1}/M$ when
there are $M$ active nodes at time $t$,
\begin{equation}
\label{Eq3}
b(M)=E\Big(\frac{x_{t+1}}{M}|x_{t}=M\Big).
\end{equation}
It is clear from definition of $b$ that for a given value of
$x_{t}=M$, if $b(M)>1$ ($b(M)<1$) an average increase (decrease)
in activity is expected in the next time step. Therefore, $b(M)$
can provide important statistical information about the behavior
of the system for the entire range of possible values of $x_{t}$.

Galton-Watson theory of branching process holds that a branching
ratio less, equal or larger than one respectively bespeaks
sub-critical, critical and super-critical phases in a system
\cite{A}. But, the activity dependent branching ratio, due to its
variability with respect to activity $x$, is different from the
branching ratio defined in the Galton-Watson process and therefore
provides much more information about the dynamics of a system. We
thus consider criticality with regard to the activity dependent
branching ratio. We define a system as critical if there exists a
range $R$ ($M\in R$) which is accessible by the long term dynamics
of the system and exhibits two characteristics: (i) the value of
the activity dependent branching ratio must be equal to one over
$R$ in the thermodynamic limit, i.e. $\lim_{N\to \infty}b(M\in
R)=1$, and (ii) $R$ must go to infinity as $N\rightarrow \infty$.
Condition (i) has to do with critical systems being (on the
average) unpredictable. Condition (ii) has to do with lack of characteristic scale for a critical systems in the thermodynamic limit. We note that $b(M)$ has been used to ascertain criticality in a wide range of systems including sandpile models of SOC or solar flares \cite{MSP} as well as neural networks \cite{AMIN}.

We begin by employing a mean-field approach in order to calculate
the activity dependent branching ratio, analytically. If there are
$M$ active nodes at a time $t$, then at time $t+1$ every node will
receive an input with the weight of $\frac{M\langle k\rangle}{N}$,
and if a node is quiescent at time $t$ will be activated with
probability of $f\big(\frac{M\langle k\rangle \sigma }{N}\big)$.
The largest eigenvalue of the connection matrix is $\lambda=\sigma
\langle k \rangle$ and the activation probability can thus be
written as $f(\frac{M\lambda}{N})$. Since $\frac{M\lambda}{N}\geq
0$, two situations are possible: (a) $0\leq \frac{M\lambda}{N}\leq
1$: in this case $f(\frac{M\lambda}{N})=\frac{M\lambda}{N}$, and
the probability of having exactly $x_{t+1}=z$ active nodes at
$t+1$ when there are $x_{t}=M$ active nodes at time $t$ can be
approximated as a binomial probability function, i.e. due to the
refractory period of one time step, there are $N-M$ nodes that can
be activated with probability $\frac{M\lambda}{N}$. (b)
$\frac{M\lambda}{N}> 1$: in this case $f(\frac{M\lambda}{N})=1$
and every quiescent node that receives input in a time step will
be activated in the next time step.  Therefore, the conditional
probability is obtained as:
\begin{equation}
\label{Eq4}
 P\Big(x_{t+1}=z|x_{t}=M\Big)=
\begin{cases}
{N-M \choose z} (\frac{M\lambda}{N})^{z}(1-\frac{M\lambda}{N})^{N-M-z} \hskip60pt \frac{M\lambda}{N}< 1\\
\\
\delta_{z,N-M} \hskip148pt \frac{M\lambda}{N}\geq 1
\end{cases}
\end{equation}
where the Kronecker delta indicates that all quiescent nodes will
on the average be activated when $M\lambda/N>1$. We can therefore
calculate the activity dependent branching ratio as:
\begin{equation}
\label{Eq5}
b(M)={\frac{E\Big(x_{t+1}|x_{t}=M\Big)}{M}}=
\begin{cases}
 \lambda-\frac{\lambda}{N}M \hskip35pt M<\frac{N}{\lambda}\\
 \frac{N}{M}-1 \hskip45pt M\geq\frac{N}{\lambda}\\
 \end{cases}
\end{equation}

\begin{figure}[!htbp]
\centering
      \includegraphics[width=0.96\linewidth]{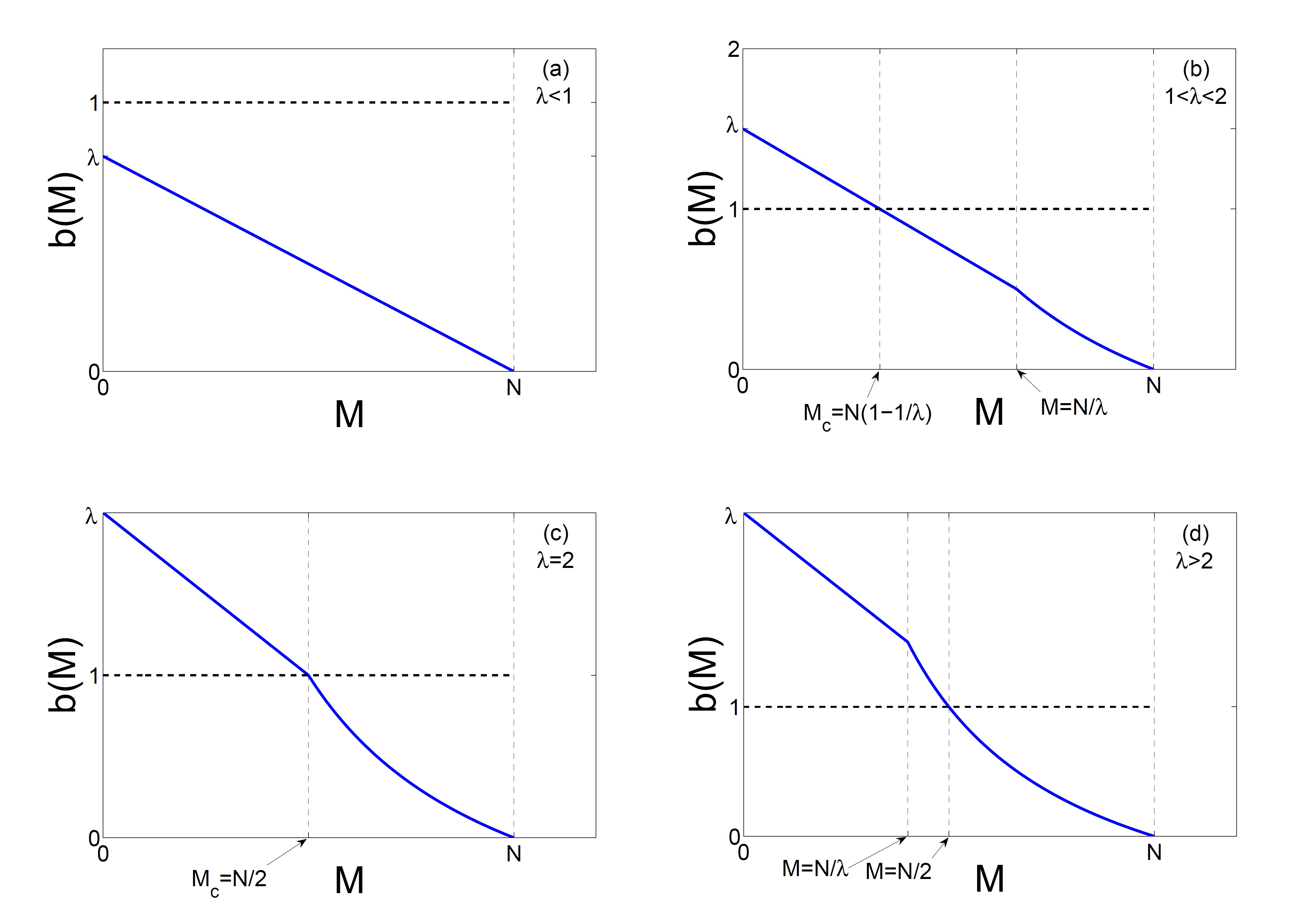}
    \caption{Activity dependent branching ratio for different parameter regimes of (a) $\lambda<1$, (b) $1<\lambda<2$, (c) $\lambda=2$ and (d) $\lambda>2$.}
    \label{Fig1}
\end{figure}

Having calculated the function $b(M)$, we can present a mean-field
analysis of the behavior of the system. As is clear from Eq.
(\ref{Eq5}) the behavior crucially depends on the value of
$\lambda$. Note that $b(M)$ is a piecewise differentiable
decreasing function of $M$, which has a linear as well as a
nonlinear regime and goes to zero at $M=N$, see Fig. \ref{Fig1}.
For $\lambda<1$ (see Fig. \ref{Fig1}(a)), we have
$M<\frac{N}{\lambda}$ and we are in the linear branch of
$b(M)=\lambda-\frac{\lambda}{N}M$ which is always less than one.
This indicates that the activity of the system will on average
decrease until reaching the fixed point of $x_{t}=x^{*}=0$ for all
initial perturbations. This is the stable sub-critical regime for
which $b(M)<1, \forall  M$. Note that, when $\lambda\to 1$,
$b(M)|_{M=x^{*}=0}\to1$ and the stable fixed point is expected to
exhibit critical behavior. Also note that in the $N \rightarrow
\infty $ limit, $b(M)=\lambda, \forall M$ for $\lambda \leq 1$
which indicates subcritical behavior for $\lambda < 1$ \emph{and}
critical behavior for $\lambda = 1$.

For $1<\lambda<2$ (see Fig. \ref{Fig1}(b)), the line $b(M)=1$ will
intersect $b(M)$ in the linear regime at
$M_{c}=N(1-\frac{1}{\lambda})$. Interestingly, due to the negative
slope of $b(M)$, $x^{*}=M_{c}$ will be the attractor of the
dynamics as $M<M_{c}$ ($M>M_{c}$) $b(M)$ will be larger (smaller)
than one which would indicate increased (decreased) average
activity until $x=M_{c}$ is reached where $b(M)=1$. This indicates
a stable, ceaseless ($x^{*}\neq 0$) dynamics which would exhibit
critical behavior as $b(M=M_{c})=1$. Note that for $\lambda=2$
(see Fig. \ref{Fig1}(c)), the critical attractor $M_{c}$ will
coincide with nonlinear regime of $b(M)$ at $M_{c}=\frac{N}{2}$.

As $\lambda$ is further increased, $\lambda>2$ (see Fig.
\ref{Fig1}(d)) the line $b(M)=1$ will intersect $b(M)$ in the
nonlinear regime and the simple analysis presented above will no
longer hold. However, one can simply note that for
$M>\frac{N}{\lambda}$, $E(x_{t+1}|x_{t}=M)=N-M$ (see Eq.
(\ref{Eq5})) and $E(x_{t+1}|x_{t}=N-M)=M$ which would indicate a
period-2 oscillating behavior. For the case when $x_{t}<N/2$ (i.e.
initial conditions in the linear regime), time evolution of the
system will increase $x_{t}$ as $b(M)>1$ in this regime until we
reach $M=\frac{N}{\lambda}$ after which the same periodic behavior
would occur between $M=\frac{N}{\lambda}$ and
$M=N-\frac{N}{\lambda}$. The fact that periodic behavior arises as a result of refractory period has previously been observed as in models of epidemic spreading \cite{Smith,Hethcote}. However, the case of refractory period larger than one presents an interesting case study, the details of which is presented in the Appendix.

\begin{figure}[!htbp]
\centering
    \includegraphics[width=0.6\linewidth]{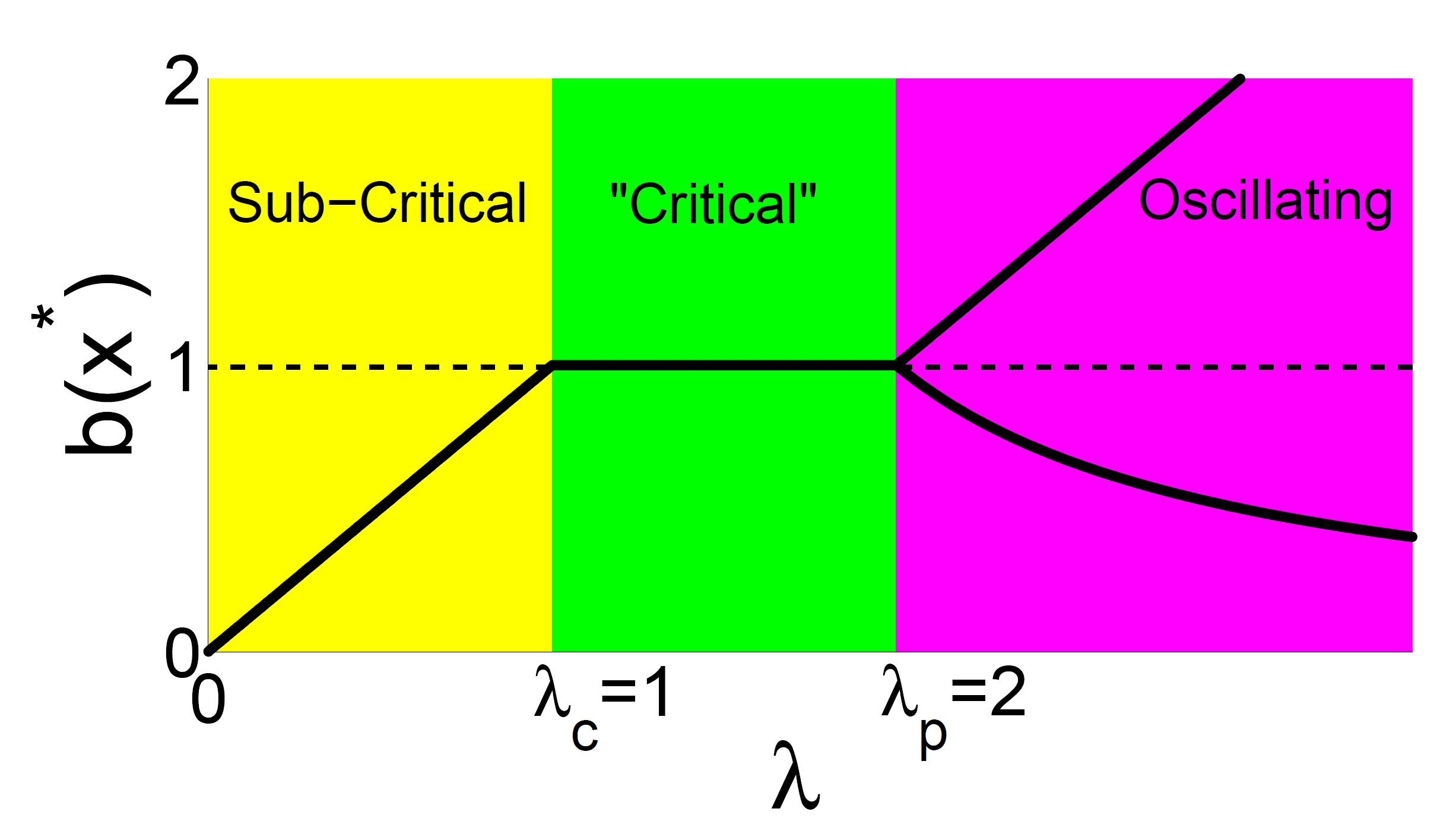}
    \caption{(Color online) Activity dependent branching ratio evaluated at the attractors ($x^{*}$) of the dynamics as a function of $\lambda$. See text for details.}
    \label{Fig2}
\end{figure}

Fig. \ref{Fig2} summarizes the mean-field behavior of the system.
Three parameter regimes are presented. For $0<\lambda<1$ the
activity dependent branching ratio at the stable attractor of the
dynamics is equal to $b(x^{*})=\lambda<1$ and the system is sub
critical. For $1.0\leq\lambda\leq2.0$ the activity dependent
branching ratio is $b(x^{*})=1.0$ at the stable attractor of the
dynamics and criticality is expected. For $\lambda>2$ we have two
values for the branching function and the dynamics jumps back and
forth between these two values. $\lambda_{c}=1$ and
$\lambda_{p}=2$ are the bifurcation points where the system
changes behavior. while the value of $\lambda_{c}$ is independent of refractory period, the value of $\lambda_{p}$ ($>1$) depends on the choice of refractory period, see Appendix.

\subsection*{Numerical calculation of b(M)}

The above results portray the general behavior of the system in
mean-field approximation where fluctuations have been ignored.
However, as is well-known fluctuations are very important and tend
to dominate system behavior in the critical regime. We therefore
propose to study our system by extensive numerical simulations for
$N=1 \times 10^{4}$ up to $8 \times 10^{4}$ and $q=0.01$ in the
range of $0.9\leq\lambda<2$ with particular focus on the critical
behavior of the system. All activities are initiated by choosing a
random site $i$ and setting $A_{i}(t=0)=1$  and following the
ensuing dynamics according to Eqs. (\ref{Eq1},\ref{Eq2}). $x_t$ is
recorded for long times from which we can easily calculate $b(M)$
numerically.

\begin{figure}[!htbp]
\centering
      \includegraphics[width=0.96\linewidth]{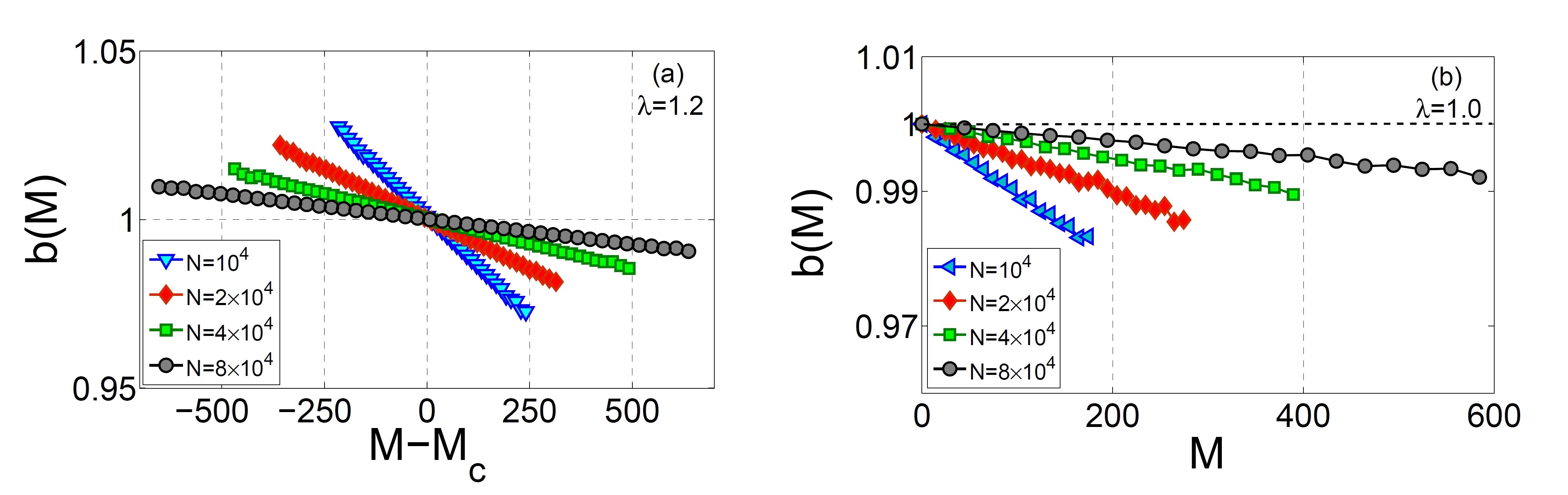}
    \caption{(Color online) Activity dependent branching ratio for (a) $\lambda=1.2$, (b) $\lambda=1$ and different values
    of $N=10^{4},2\times10^{4},4\times10^{4},8\times10^{4}$}
    \label{Fig3}
\end{figure}

Fig. \ref{Fig3} shows our results for $b(M)$ for various system
sizes and $\lambda=1.2$ as well as $\lambda=1.0$, where
criticality is expected. In both cases, our results clearly show a
linear behavior in accordance with Eq. (\ref{Eq5}), where
$b(M)=\lambda-\frac{\lambda}{N}M$ provides a prefect fit to the
data. Note that as $N$ is increased the range of system's activity
increases as the slope ($\frac{\lambda}{N}$) goes to zero.
Therefore, one can expect that in the large system size limit
$b(M)\to 1$ for all accessible $M$ indicating a critical behavior.
We have also checked various other values of $\lambda$ and have
observed similar behavior to that of $\lambda=1.2$ (above) for the
range of $1<\lambda<2$ (not shown).

 \begin{figure}[!htbp]
\centering
    \includegraphics[width=0.6\linewidth]{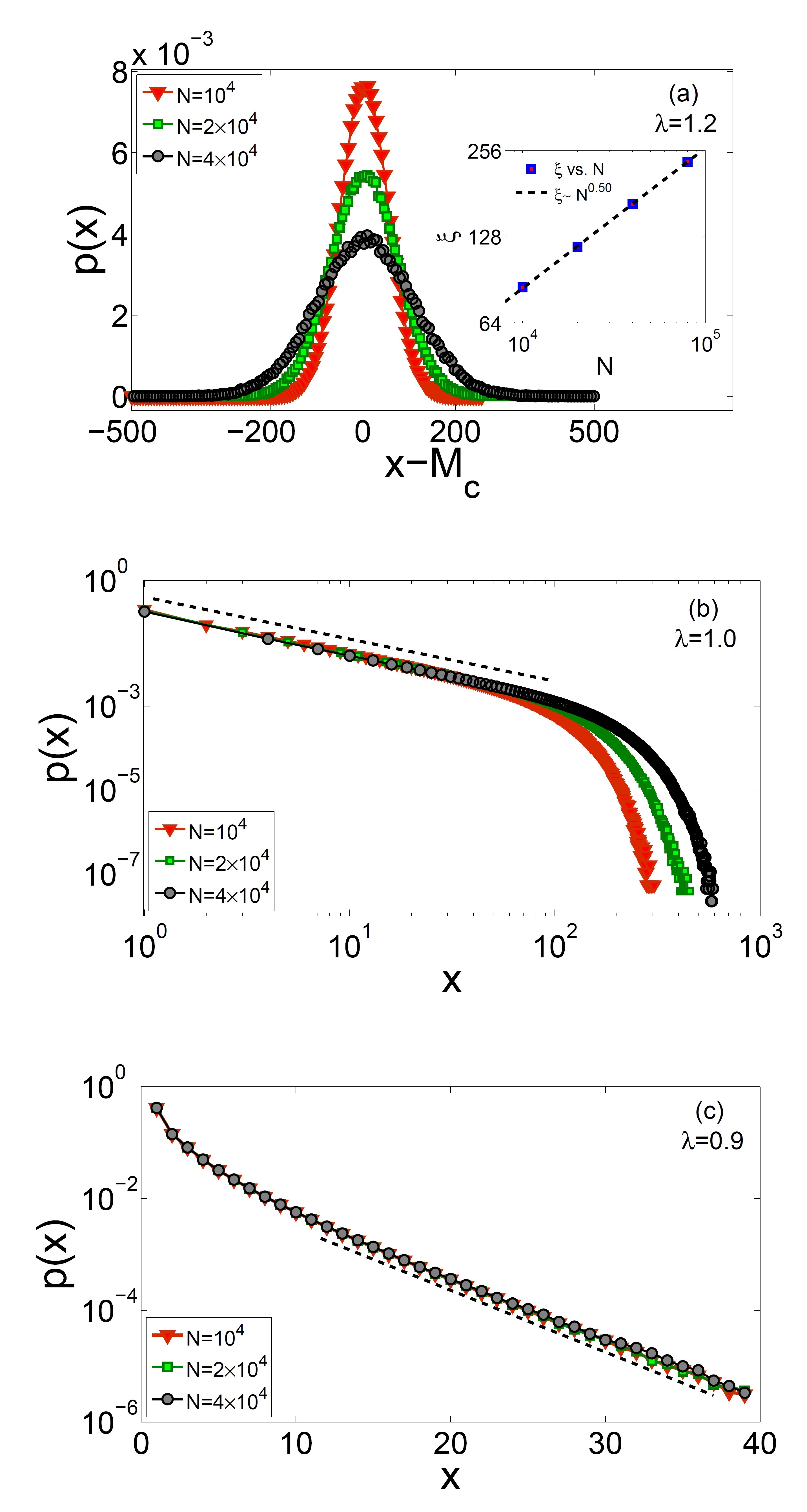}
    \caption{(Color online) Probability distribution function of aggregate activity ($p(x)$) as a function of $x$ for different values of $N$.
    Panel (a) is a linear plot showing a Gaussian function for $\lambda=1.2$ with a maximum at $x=x_{c}=M_{c}$.
    Inset is a log-log plot of the standard deviation $\xi$ versus $N$. Panel (b) is a log-log plot showing power-law behavior for $\lambda=1.0$,
    and panel (c) is a semi-log plot with $\lambda=0.9$ showing exponential behavior. }
    \label{Fig4}
\end{figure}

\subsection*{Avalanche statistics}
In order to better understand the behavior of fluctuations about
the attractors $x^{*}$, we have plotted the probability
distribution function of system's activity as $p(x)$ in Fig.
\ref{Fig4} for $\lambda=1.2$, $1.0$, and $0.9$. For $\lambda=1.2$
(Fig. \ref{Fig4}(a)) we observe a Gaussian behavior which peaks
exactly at $x=M_{c}=N(1-\frac{1}{\lambda})$. It is interesting to
note that our results indicate that the width of the Gaussian
increases with the system size as $\xi\sim N^{0.5}$ (see inset of
Fig. \ref{Fig4}(a)), in accordance with the central limit theorem.
For $\lambda=1.0$ (Fig. \ref{Fig4}(b)), we observe a distinctly
different behavior as $p(x)$ displays a power-law behavior with
system size dependent cutoff. It is, however, important to note
that $p(x)$ is maximized at $x=0$ as indicated by our mean-field
analysis. In Fig. \ref{Fig4}(c), we plot the same results for
$\lambda=0.9$ for various system sizes on a log-linear plot, all
of which coincide on the same curve. We therefore conclude that
$p(x)$ displays an exponential behavior with a scale which is
system size \textit{independent}. Again, the attractor $x=0$
appears as the most probable state, however, the size independent
scale in (c) as opposed to size dependent scale in part (b) (and
even (a)) is the distinction between sub-critical and critical
systems.

\begin{figure}[!htbp]
\centering
    \includegraphics[width=0.6\linewidth]{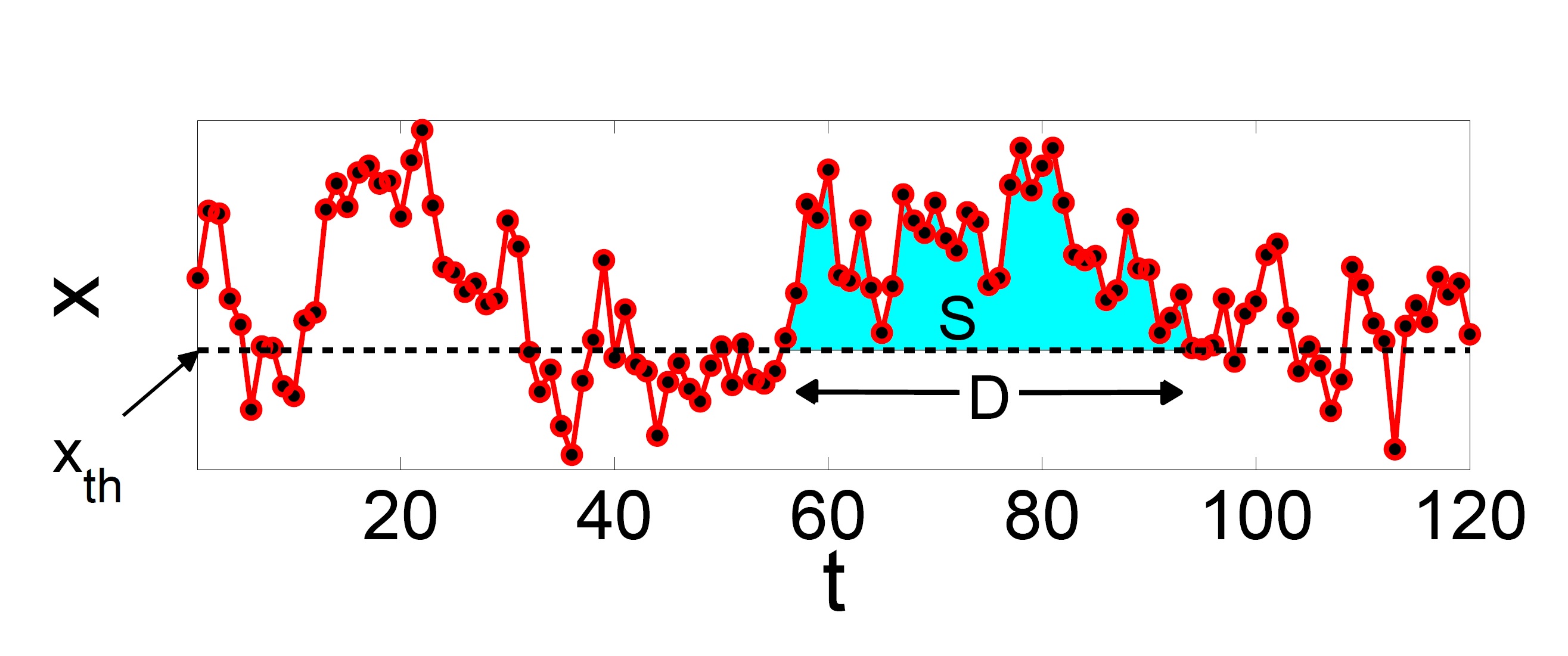}
    \caption{(Color online) An avalanche is defined as excursion of the aggregate activity of nodes above a threshold value $x_{th}$,
    $D$ is the defined as the duration and the integral between $x_{t}$ and $x_{th}$ (the colored area) as the size ($S$) of the avalanche. }
    \label{Fig5}
\end{figure}

As we have defined a critical system based on the behavior of
$b(M)$ (see above), we have shown that our system exhibits
critical behavior in the range $1\leq\lambda<2$. However,
distinctly different behavior is observed for $P(x)$ as $M_{c}=0$
(Fig. \ref{Fig4}(b)) changes to $M_{c}\neq0$ (Fig. \ref{Fig4}(a)). To
better understand the critical behavior of the system we now turn
our attention to avalanches. In the case of $\lambda=1.0$ for
which the stable attractor of the dynamics is $x^{*}=0$, the
avalanches are well defined as the activity between two
stabilities initiated by an external perturbation. For other
values of $1<\lambda<2$ over which the system exhibits
self-sustaining behavior we define the avalanches (see Fig.
\ref{Fig5}), as the continuous aggregate activity of nodes above a
threshold value $x_{th}$. The number of time steps of an excursion
above $x_{th}$ is defined as duration (D) and the summation
$\sum_{D} x_{t}-x_{th}$ as the size ($S$) of an avalanche.

Probability distribution function of size and duration of
avalanches ($P(y)$ , $y\in \{S,D\}$) are calculated for systems
with different $N$ over the parameter regime $1\leq\lambda\leq2$.
In the case of criticality, these probability distribution
functions are expected to exhibit power-law behavior with a cutoff
which is an increasing function of $N$. The usual scaling ansatz
for such a behavior is $P(y)\sim y^{-\tau_{y}}
g_{y}(y/N^{\beta_{y}})$, where $g_{y}$ is a universal cutoff
function that is identical for different system sizes. $\tau_{y}$
is the critical exponent and $\beta_{y}$ is referred to as the
finite-size scaling exponent. When criticality holds, if we
rescale $y\to y/N^{\beta_{y}}$ and $P(y)\to y^{\tau_{y}}P(y)$ then
the plots of rescaled variables must collapse into one universal
curve for the correct values of $\tau_{y}$ and $\beta_{y}$
\cite{P}.

\begin{figure}[!htbp]
\centering
    \includegraphics[width=0.6\linewidth]{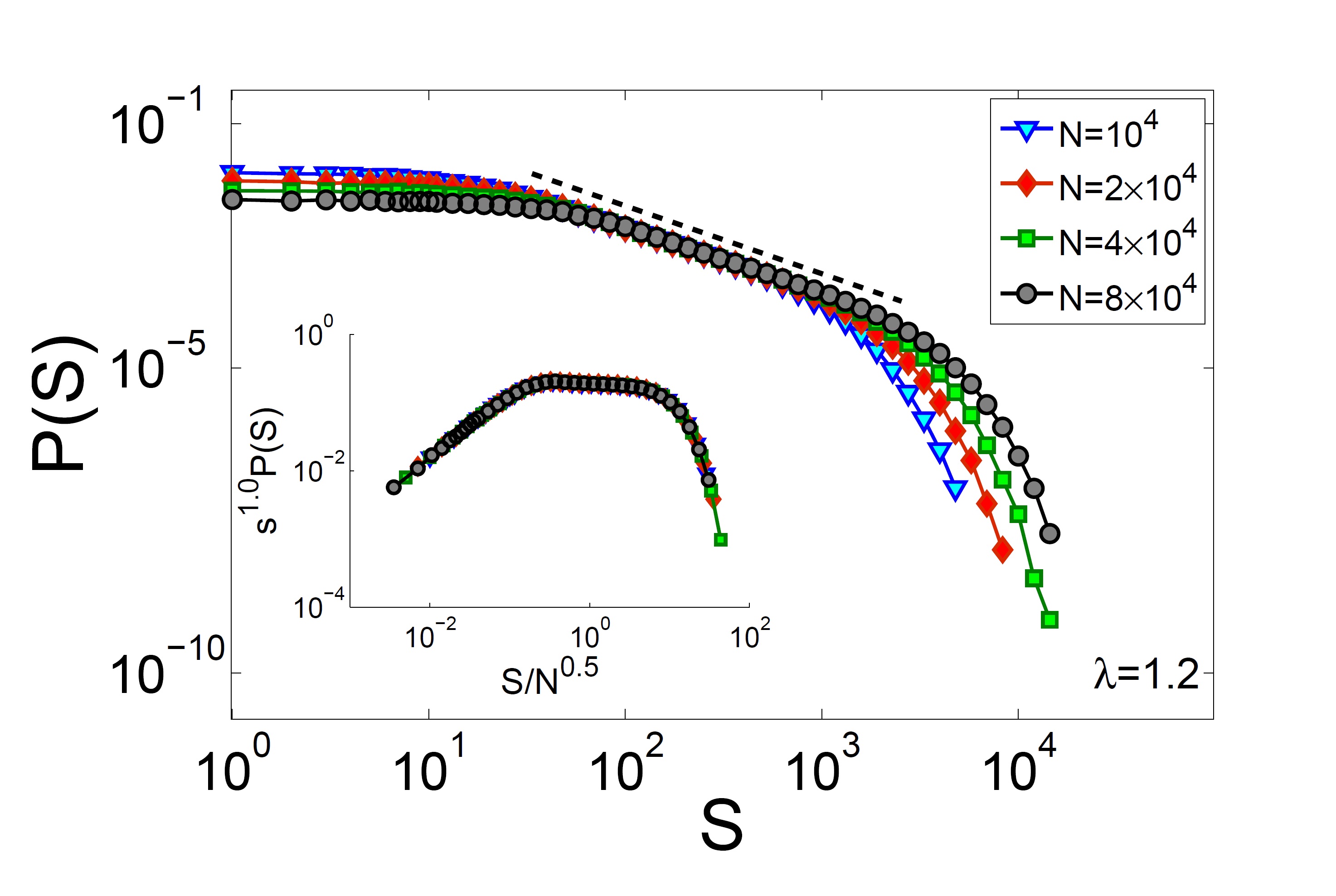}
    \caption{(Color online) Main panel: probability distribution functions of avalanche sizes for systems with $\lambda=1.2$
    and different values of $N$. Inset panel: plots of rescaled data, collapsed into one universal curve with $\tau_{S}=1.0$ and $\beta_{S}=0.5$.}
    \label{Fig6}
\end{figure}

Probability distribution functions of $S$ are plotted in the main
panel of Fig. \ref{Fig6} for systems with $\lambda=1.2$ and
different sizes. It is clearly seen that the plots exhibit a
power-law region and a cutoff that increases by the system size.
Inset panel of Fig. \ref{Fig6} shows collapse of the rescaled data
of the main panel with exponents $\tau_{S}=1.00\pm0.01$ and
$\beta_{S}=0.50\pm0.01$. It must be noted that our numerical
analysis show that different choices of $x_{th}$ does not change
the values of $\tau_{S}$ and $\beta_{S}$, and we choose
$x_{th}=x_{c}$ where we have better statistics for avalanche
distribution functions. Our numerical analysis for other values of
$1<\lambda<2$ indicate that the same values of
$\tau_{S}=1.00\pm0.01$ and $\beta_{S}=0.50\pm0.01$ are also
obtained.

\begin{figure}[!htbp]
\centering
    \includegraphics[width=0.6\linewidth]{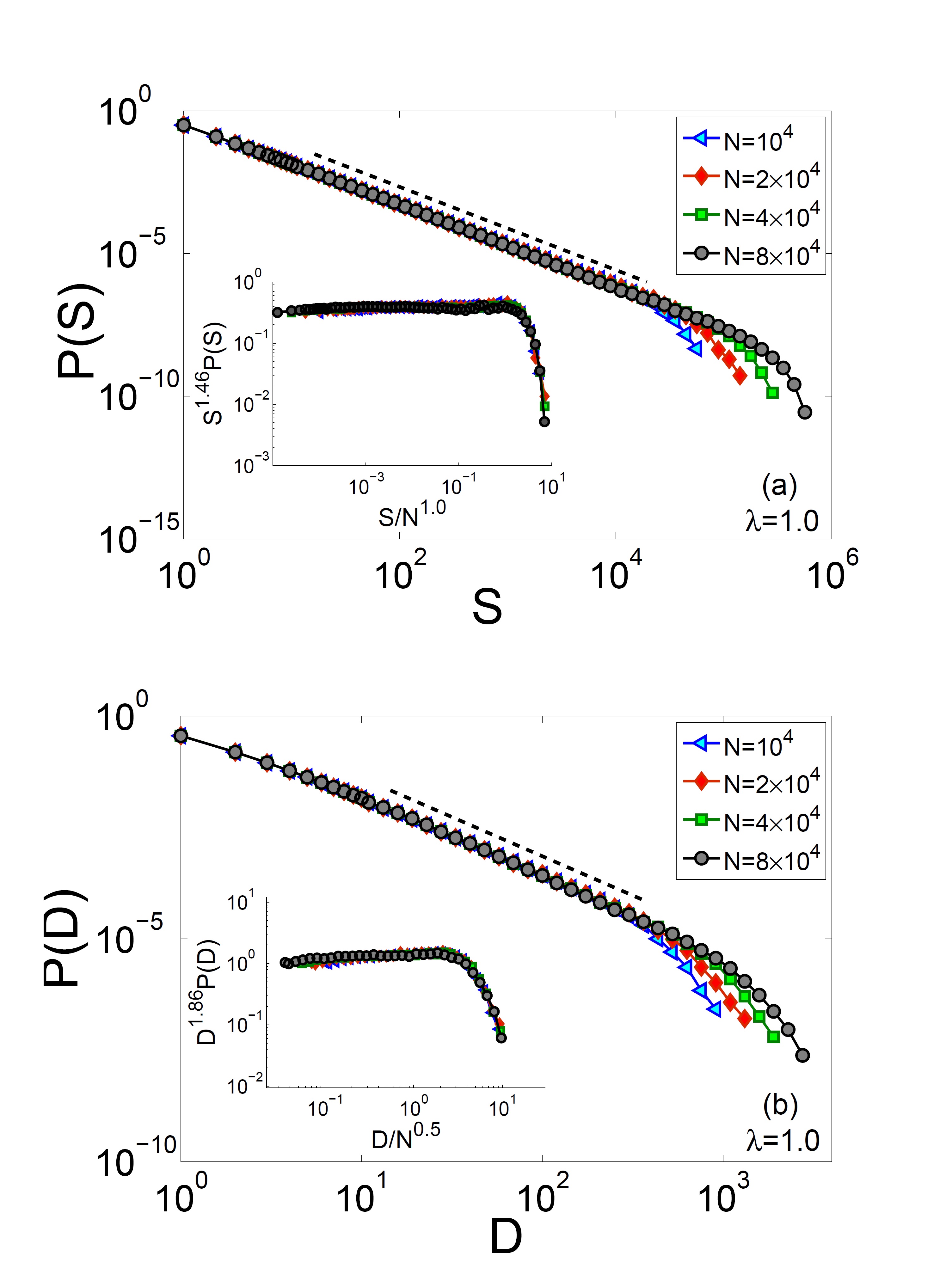}
    \caption{(Color online) Main panels: probability distribution functions of (a) avalanche sizes and (b) avalanche durations
    for systems with $\lambda=1.0$ and different values of $N$. Inset panel: plots of rescaled data, collapsed into one universal curve
    with $\tau_{S}=1.46$, $\beta_{S}=1.0$, $\tau_{D}=1.86$, and $\beta_{D}=0.5$.}
    \label{Fig7}
\end{figure}

However, the critical behavior obtained for $\lambda=1.0$ is
somewhat different. As shown in Fig. \ref{Fig7}(a), we obtain
$\tau_{S}=1.46\pm0.02$ and $\beta_{S}=1.00\pm0.03$ which are
indication of a different universality class for $\lambda=1.0$,
where the critical exponent is close to the critical branching
process, i.e. $\tau_{S}=3/2$. More importantly, however, our study
of finite-size scaling of avalanche sizes, in addition to $b(M)
\rightarrow 1$ presented earlier, provide firm evidence for
critical behavior of the model in the range of $1\leq \lambda <2$.
This could potentially provide an explanation to a wide range of
criticality observed in neuronal systems, without any apparent
tuning of parameters.

We have also calculated probability distribution functions of
avalanche durations ($P(D)$). In Fig. \ref{Fig7}(b), we have shown reslults
for the $\lambda=1.0$ case, where we obtain $\tau_{D}=1.86\pm0.02$
and $\beta_{D}=0.50\pm0.01$ again close to the exponent
$\tau_{D}=2.0$ of the critical branching process. However, the
model exhibits an unusual scaling behavior for $D$ in
$1<\lambda<2$ range.  For system sizes we have been able to study
(i.e up to $N=8 \times 10^{4}$), probability distribution
functions of avalanche durations \emph{do} exhibit a power-law
behavior. However, no appreciable increase is observed in the
cutoff for the present system sizes.  This behavior could possibly
happen if the finite-size exponent $\beta_{D}$ is so small that
the cutoff function does not change considerably for the system
sizes we have considered here.

\begin{figure}[!htbp]
\centering
    \includegraphics[width=0.6\linewidth]{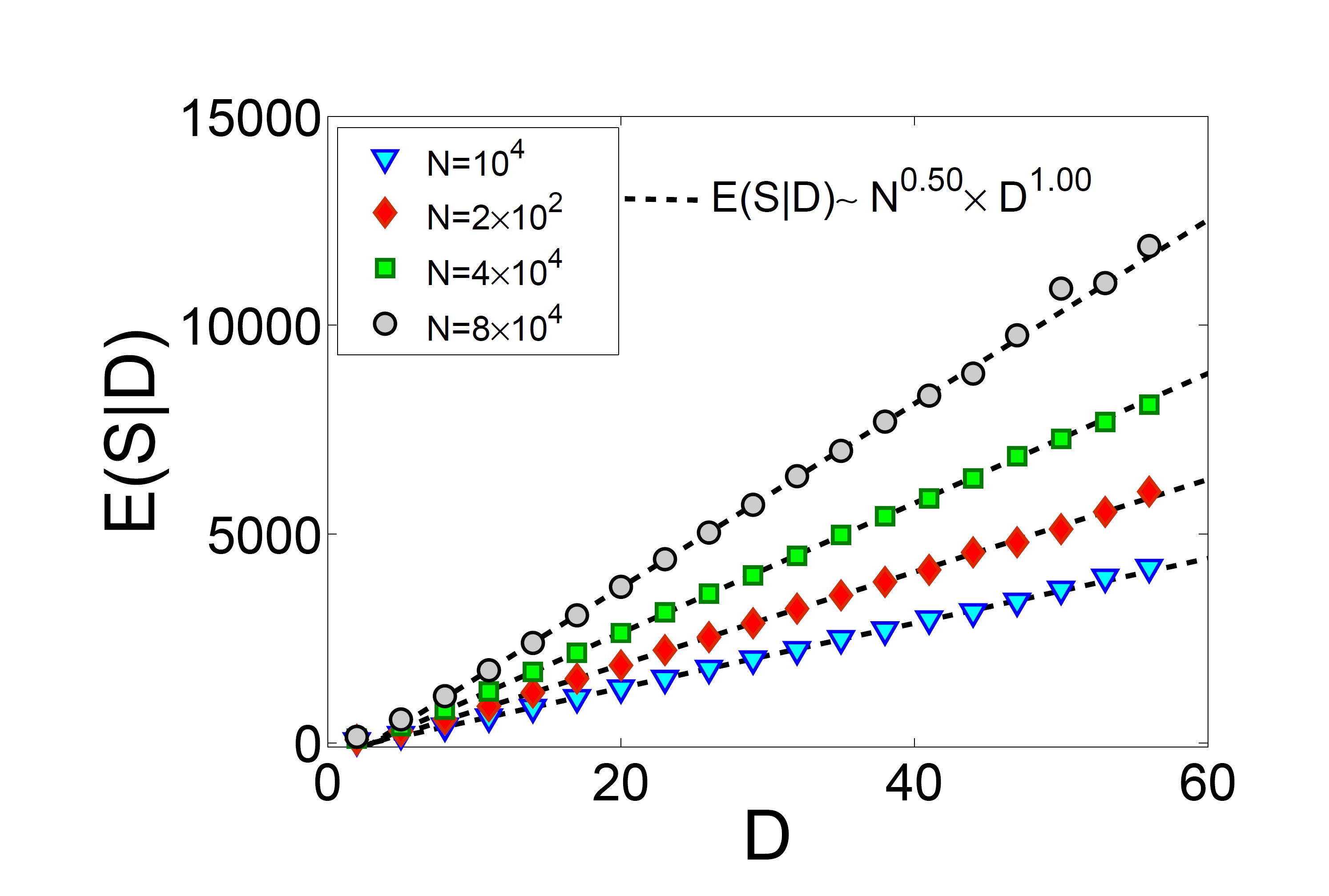}
    \caption{(Color online) Plots of $E(S|D)$ versus $D$ for systems with $\lambda=1.2$ and different
    values of $N$. Dashed lines are plots of $N^{\alpha}\times D^{\gamma}$ fitted to the data with $\alpha=0.5$ and $\gamma=1.0$.}
    \label{Fig8}
\end{figure}

To shed light on such a behavior, we consider a scaling anzats
that relates the size and duration of avalanches as:
\begin{equation}
\label{Eq6}
E(S|D)\sim N^{\alpha}D^{\gamma}
\end{equation}
in which $E(S|D)$ is the expectation value of $S$ when $D$ is
given.  As seen in Fig. \ref{Fig8}, $E(S|D)$ is a linear function
of $D$ for a given system size. Careful regression analysis shows
that $\alpha=0.50\pm0.01$ and $\gamma=1.00\pm0.01$. On the other hand, from
the above numerical analysis, we know that the maximum value of
avalanche sizes scales as $S_{max}\sim N^{\beta_{S}}$. Due to the
linear relation between $E(S|D)$ and $D$ we can write
$E(S|D_{max})=S_{max}$. Using Eq. (\ref{Eq6}) we write
$S_{max}\sim N^{\alpha}D_{max}$ which leads to $D_{max}\sim
N^{\beta_{S}-\alpha}$ and therefore gives
$\beta_{D}=\beta_{S}-\alpha=0.00\pm0.02$. The fact that
$\beta_{D}\approx0$ indicates why we do not observe finite size
scaling for avalanche durations despite the fact that we observe a
power-law behavior for $P(D)$ in a limited range of data. This is
an interesting case whose full understanding requires further
investigation with much larger system sizes than studied here.

\subsection*{Sensitive dependence to perturbations}
\begin{figure}[!htbp]
\centering
     \includegraphics[width=0.6\linewidth]{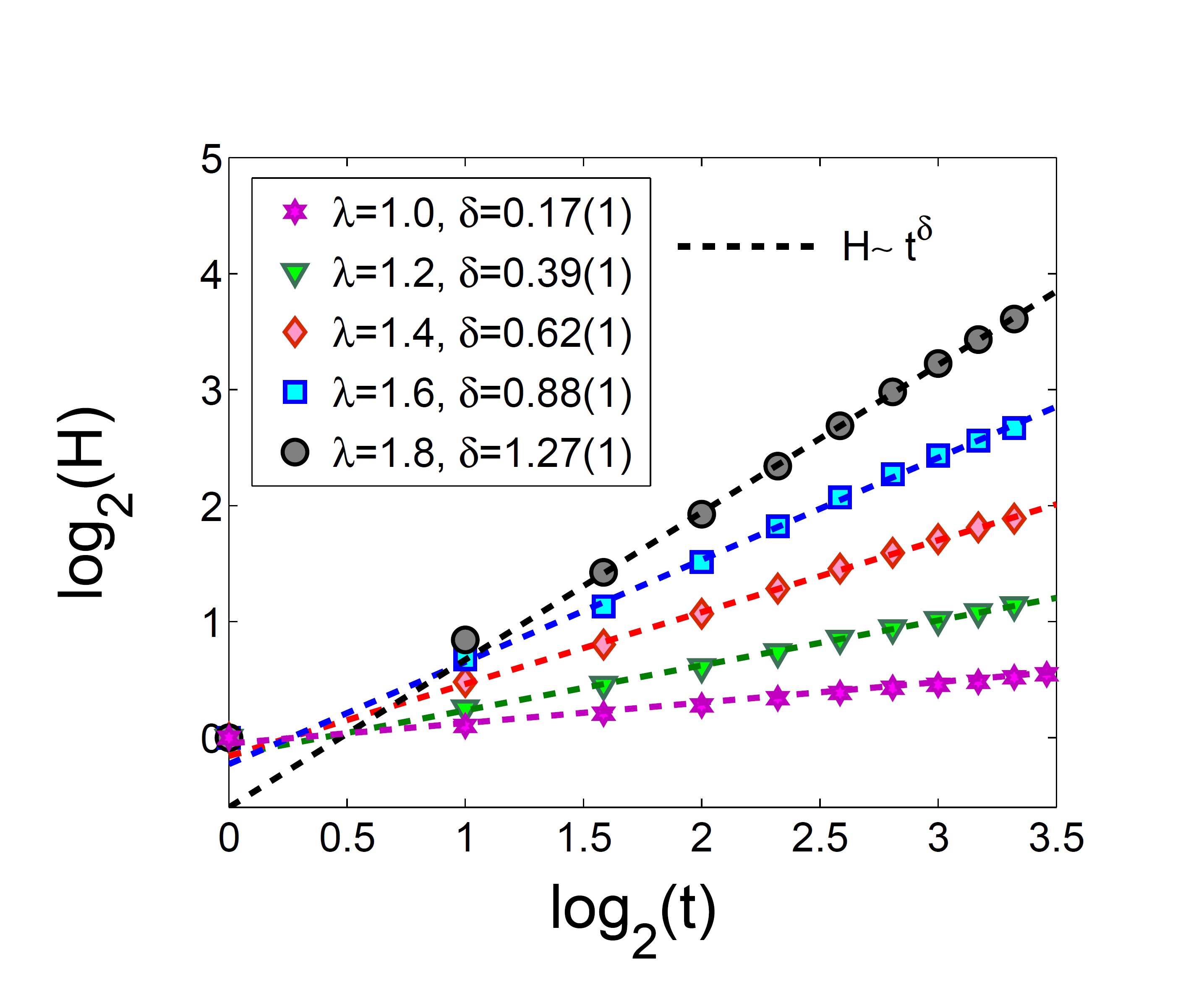}
    \caption{(Color online) Short term evolution of the Hamming distance averaged over $2000$ realizations for systems with one refractory time step, $N=40000$, and different values of $\lambda=1.0, 1.2, 1.4, 1.6, 1.8$. Linear fits are indications of power-law behavior.   }
    \label{Fig9}
\end{figure}

An interesting property of critical dynamical systems, such as self-organized critical models, is that short-term evolution of perturbations is a power-law function of time and the system exhibits power-law sensitivity to initial conditions \cite{Pinho}. In order to present another evidence for criticality of the system in the parameter regime $1\leq\lambda<\lambda_{p}$, we test this behavior in our model. In order to do that we consider a system in this parameter regime and run the simulation until the system reaches its \lq\lq critical'' state. At this point we pause the simulation for a moment. We have the activity vector $\textbf{A(}t\textbf{)}=\{A_{1}(t), A_{2}(t), A_{3}(t), ... , A_{N}(t)\}$ in an $N$ dimensional space and make a copy of the system $\textbf{A}^\prime$. Then, we introduce a small perturbation to $\textbf{A}^\prime$ by changing a few randomly chosen elements ($A^{\prime}_{i}$) from one to zero or vice versa. The difference between two systems, which is a distance in the N dimensional space, is defined as the Hamming distance ($H$) between $\textbf{A}$ and  $\textbf{A}^\prime$ which is
\begin{equation}
\label{Eqq8}
H(t)=\sum_{1}^{N} (A_{i}(t)-A^{\prime}_{i}(t))^2
\end{equation}
Now, we can study the short-term evolution of $H(t)$. It must be noted that we used the same random seed for simulating both systems. In order to have firm results we need to do an ensemble averaging, therefore, we have done the above process for $2000$ realizations. Time evolution of the ensemble-averaged Hamming distance is plotted in Fig. \ref{Fig9} for systems with refractory period of one time step, $N=40000$ and different values of $\lambda=1.0, 1.2, 1.4, 1.6, 1.8$. It is observed that the system exhibits power-law sensitivity to initial conditions
\begin{equation}
\label{Eqq8}
H(t)\sim t^{\delta}
\end{equation}
This provides yet another evidence for criticality of the system in the parameter range of $1\leq \lambda<\lambda_{p}$. Note that we have also included $\lambda_{c}=1.0$ where criticality is well established. The exponent $\delta$ is an increasing function of $\lambda$ indicating more chaotic behavior.

\section*{Concluding Remarks}

Motivated by the fact that scaling behavior is observed in systems
composed of interacting excitable nodes, and that excitable nodes
can exhibit refractory period, as in neuronal systems,  we have
studied a simple model of excitable nodes on a random directed
graph in the presence of refractory period. The behavior of the
model without refractory period has been well-studied previously.

We find that in the presence of refractory period the behavior of
system undergoes dramatic changes and different dynamical regimes
become accessible for the system. A sub-critical regime for
$\lambda<1.0$. A standard critical behavior with scaling and
power-law behavior of avalanche sizes and durations, similar to
the critical branching process, for $\lambda=1$. An important
regime with stable self-sustaining dynamics with interesting
scaling behavior for $1<\lambda<\lambda_{p}$ where activity dependent
branching ratio goes to one in the thermodynamic limit and the
system exhibits power-law statistics in avalanche distribution
functions as well as power-law sensitivity to initial perturbations. The critical exponents associated with this new regime
are distinct from those obtained from the standard critical point
at $\lambda = 1$. Finally an oscillating regime with $\lambda>\lambda_{p}$
where the dynamics oscillates between various well-defined states is
observed. While our results highlights the importance of
refractory period in networks of excitable nodes, it can also
provide understanding for similar behavior observed in real
neuronal systems. Branching ratios equal to one, as well as power
law statistics in neuronal avalanches have been observed in a wide
variety of real neuronal systems, with no apparent tuning of a
parameter. The fact that our model exhibits similar behavior for a
wide range of parameter (i.e. $1\leq \lambda < \lambda_{p}$) is the main
result of our study. The more general case of longer refractory periods does not change our general conclusions and the details of such generalization appear in the Appendix.

In a recent work Larremore \textit{et al.} have shown that
inhibition causes ceaseless dynamics at or near the critical point
($\lambda\lesssim1$) in a similar model \cite{LSOSR}. We, on the
other hand, have also observed ceaseless stable dynamics but in a
different parameter regime of $\lambda>1$, when refractory period
is included without inhibition.  Both inhibition and refractory
period are thought to decrease the level of activity in a system.
However, they seem to lead to stable ceaseless dynamics in
networks of excitable nodes. It would be interesting to study the
effect of both these important properties simultaneously to see if
larger parameter regime with stable dynamics and scaling
(critical) behavior can be obtained.

It is generally believed that networks with small-world effect
would exhibit critical behavior with exponents associated with
critical branching process which correspond to mean-field
exponents $\tau_S = 3/2$ and $\tau_D = 2$. Here, we have observed
non-mean-field behavior with exponents which are significantly
smaller than the critical branching process for a wide range of
parameter in a small-world network of excitable nodes with
refractory period. Of course, we have also obtained similar
mean-field exponents for a particular value of $\lambda =1$.

Although neuronal dynamics have been the main motivation of our
work, other important dynamical processes such as epidemic
spreading could also have relevance to our work as refractory
period is thought to be important in such spreading processes as
well.

\section*{Numerical details}
Computer code simulations were developed in FORTRAN 90. In order to get good statistics for probability distribution functions of $x$ as well as duration and size of avalanches, for each system size ($N=10^4, 2\times10^4, 4\times10^4, 8\times10^4$) we performed simulations for 20 different realization of networks and $10^6$ time steps for each network. Random networks were made by having every two node connected with a probability of $q=0.01$.

\section*{Acknowledgements}

Support of Iranian National Elites Foundation, Shiraz University
Research Council and Institute for Advanced Studies in Basic
sciences (IASBS) is acknowledged. S.A.M. also acknowledges Y.
Souboti without whose support this work could not be completed.

\section*{Author contributions statement}

S.A.M. and A.M. conceived the project. S.A.M. carried out the
numerical simulations as well as analytical calculations. S.A.M,
A.M. and A.V. analyzed the results. S.A.M and A.M wrote the
manuscript. All authors reviewed and approved the manuscript.

\section*{Additional information}

Authors declare no competing financial interests.

%The corresponding author is responsible for submitting a \href{http://www.nature.com/srep/policies/index.html#competing}{competing financial interests statement} on behalf of all authors of the paper. This statement must be included in the submitted article file.

\section*{Appendix\\Dynamics of the system with arbitrary refractory period}

To analyze the behavior of system for larger refractory periods, we consider $1+r$ refractory time steps, for every node, after each excitation. Thus, we can rewrite the probability of excitation (Eq. (1) in the main text) in the more general form of
\begin{equation}
\label{Eqq1}
p\big(A_{i}(t+1)=1\big)=\prod_{T=0}^{r}\delta_{0,A_{i}(t-T)} f\big(\sum_{j=1}^{N} w_{ij}A_{j}(t)\big)
\end{equation}
We can now provide a mean-field approximation for the probability of having $x_{t+1}=z$ when $x_{t}=M, x_{t-1}=M_{1}, ..., x_{t-r}=M_{r}$ as
\begin{equation}
\label{Eqq4}
 P\Big(x_{t+1}=z|x_{t}=M,x_{t-1}=M_{1},...,x_{t-r}=M_{r}\Big)=
\begin{cases}
{N-M-\sum_{i=1}^{r}M_{i} \choose z} (\frac{M\lambda}{N})^{z}(1-\frac{M\lambda}{N})^{N-M-\sum_{i=1}^{r}M_{i}-z} \hskip40pt \frac{M\lambda}{N}< 1\\
\\
\delta_{z,N-M-\sum_{i=1}^{r}M_{i}} \hskip158pt \frac{M\lambda}{N}\geq 1
\end{cases}
\end{equation}
Next we can calculate the expectation value $E(x_{t+1}|x_{t}=M,x_{t-1}=M_{1},...,x_{t-r}=M_{r})$ which results in the activity dependent branching ratio. By setting $\Delta=\sum_{i=1}^{r}M_{i}$ we will have
\begin{equation}
\label{Eqq5}
b(M,\Delta)=
\begin{cases}
 \lambda-\frac{\lambda}{N}M-(\frac{\lambda}{N})\Delta \hskip35pt M<\frac{N}{\lambda}\\
 \\
 \frac{N-\Delta}{M}-1 \hskip75pt M\geq\frac{N}{\lambda}\\
 \end{cases}
\end{equation}
We see that $b(M)$ is again a piecewise function with respect to $M$, a linear (negative slope) and a nonlinear piece (see Fig. 1 in the main text). Clearly the value of $b$ at each time step ($t$) depends on $x_{t}$ as well as the history of the system via $\Delta$. To analyze the behavior of the system we must be careful that for each value of $M$, depending on the history of the system, there are $N-M$ different possible values for $b(M,\Delta)$  ($\Delta$ can have any integer value between $1$ and $N-M$).

For $\lambda<1$ we see that $M<\frac{N}{\lambda}$ and $b(M,\Delta)<1$ for all possible values of $M$ and $\Delta$. Therefore we expect sub-critical behavior in this range of $\lambda$. But, analysis of the system for $\lambda\geq 1$ is not that simple. In order to present a mean-filed study of the system we start with a random initial condition by giving random integer values to $x_{t=r}=M, x_{t=r-1}=M_{1}, x_{t=r-2}=M_{2}, ... ,x_{0}=M_{r}$ with the constraint $\Delta<N$, and following the dynamics using the map
\begin{equation}
\label{Eqq6}
  x_{t+1}=x_{t}\times b(x_{t},\Delta_{t})
\end{equation}
Note that this is the expected dynamics (map) according to our mean-field approximation. We consider a phase space of $(x,\Delta)$ and study the time evolution of the system in order to find possible attractors. If the dynamics exhibits only a fixed point $(x^{*},\Delta^{*})$ in which $\Delta^{*}$ must be equal to $r\times x^{*}$, then we must have $b(x^{*},\Delta^{*})=1$. Using Eq. (\ref{Eqq5}) and the assumption that $x^{*}<N/\lambda$ we can calculate $x^{*}$ to be:
\begin{equation}
\label{Eqq7}
x^{*}=\frac{N(1-1/\lambda)}{1+r}
\end{equation}
Which is consistent with the $r=0$ case obtained in the main text. Accordingly, we expect critical fluctuations about this point.

\begin{figure}[!htbp]
\centering
    \includegraphics[width=0.96\linewidth]{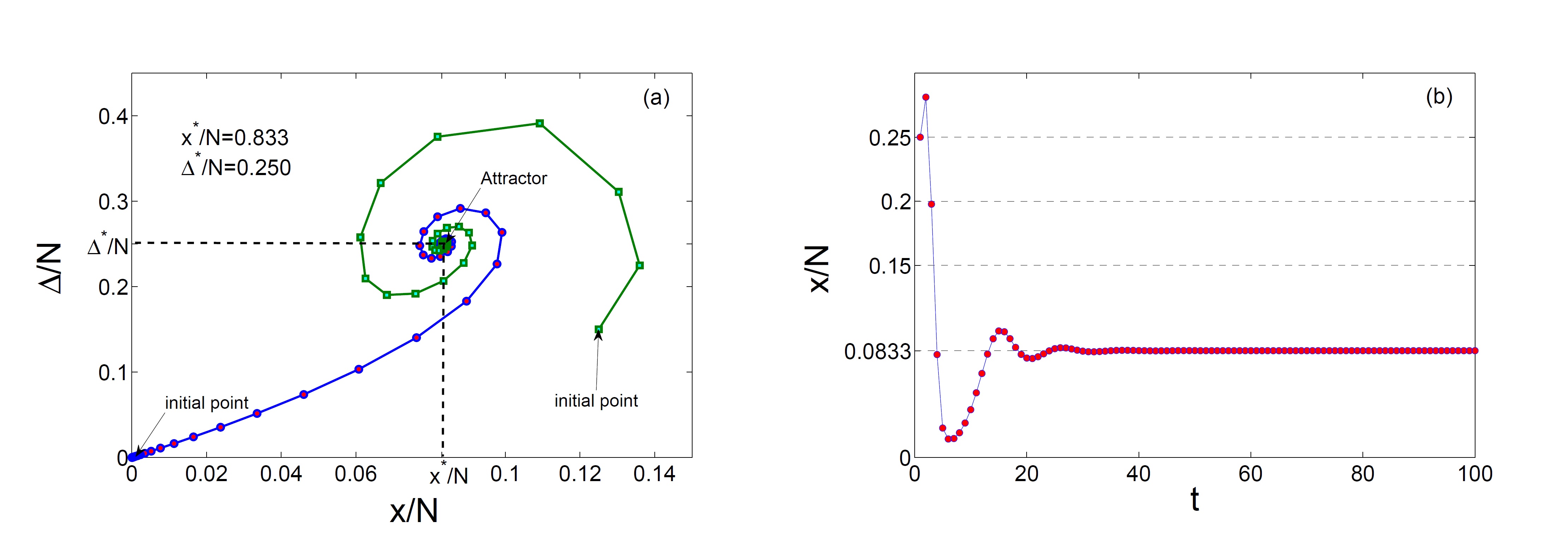}
    \caption{(Color online) Mean-Field analysis for a system with $r=3$, $\lambda=1.5$ and $N=40000$. (a) Starting from different initial conditions the dynamics is attracted to a fixed point. (b) Time evolution of $x/N$ and saturation at the attractor. }
    \label{Fig10}
\end{figure}

In order to understand the behavior of the system we numerically solved Eq. (\ref{Eqq6}) for different values of $r$ and $\lambda$. For small values of $\lambda$ we find a fixed point which is in agreement with Eq. (\ref{Eqq7}), and for larger $\lambda$ an oscillatory behavior is observed. As is shown in Fig. \ref{Fig10} (a prototypical example with small $\lambda$), the system exhibits a fixed point for $\lambda=1.5$, $r=3$ and $N=40000$. Fig. \ref{Fig10}(a) shows the trajectories in the phase space that end at the attractor of the dynamics. The values of $x^{*}=0.833\times N$ and $\Delta^{*}=0.250\times N$ are in agreement with Eq. (\ref{Eqq7}). Fig. \ref{Fig10}(b) shows the time evolution of $x$ and its saturation at $x^{*}$.

\begin{figure}[!htbp]
\centering
      \includegraphics[width=0.9\linewidth]{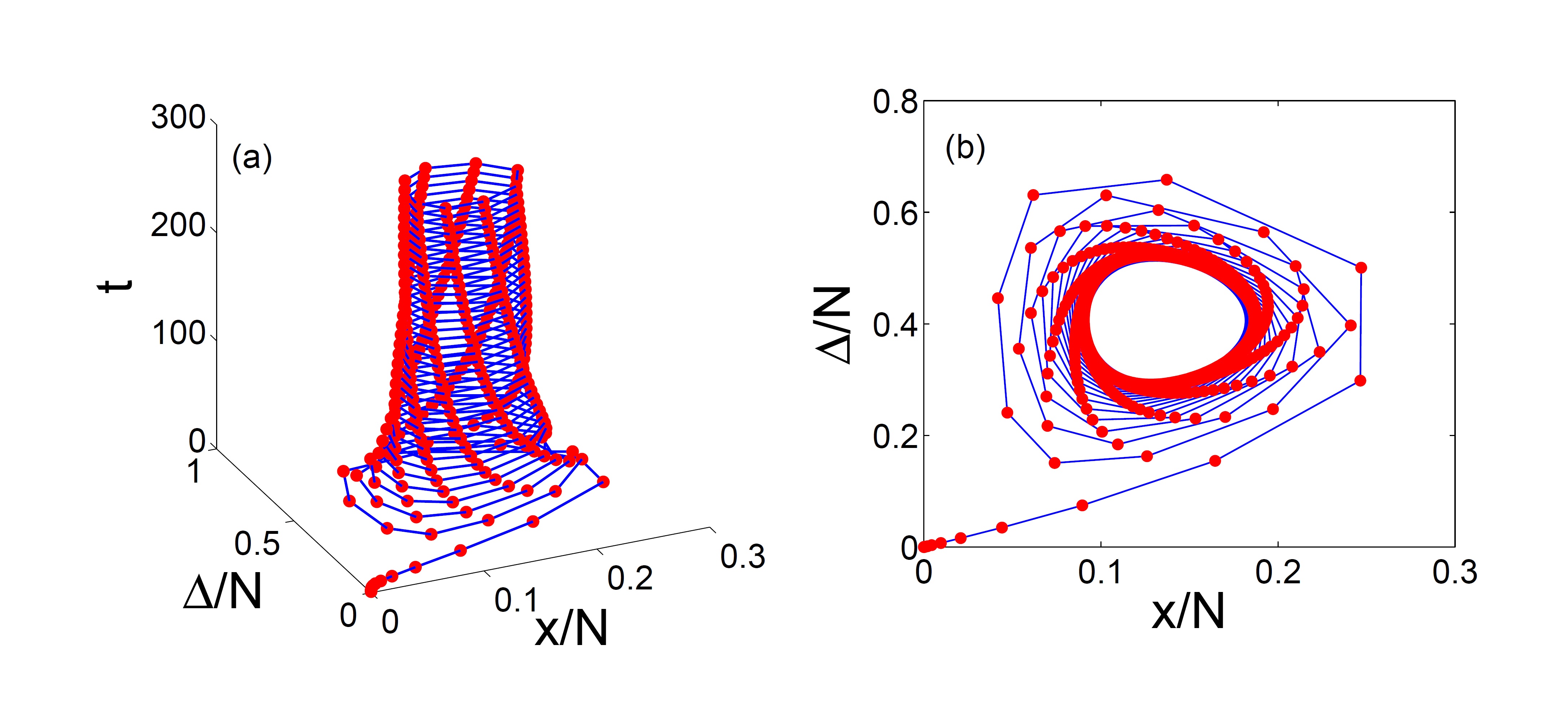}
    \caption{(Color online) Mean-field analysis of a system with $r=3$, $\lambda=2.2$ and $N=40000$. (a) Time evolution of the dynamics in the phase space. Dynamics repeatedly jumps between 8 points in the phase space that are rotating. (b) A two dimensional view of panel (a), the egg-shaped attractor of the dynamics in the phase space where the 8 points rotate on.}
    \label{Fig11}
\end{figure}

\begin{figure}[!htbp]
\centering
    \includegraphics[width=0.9\linewidth]{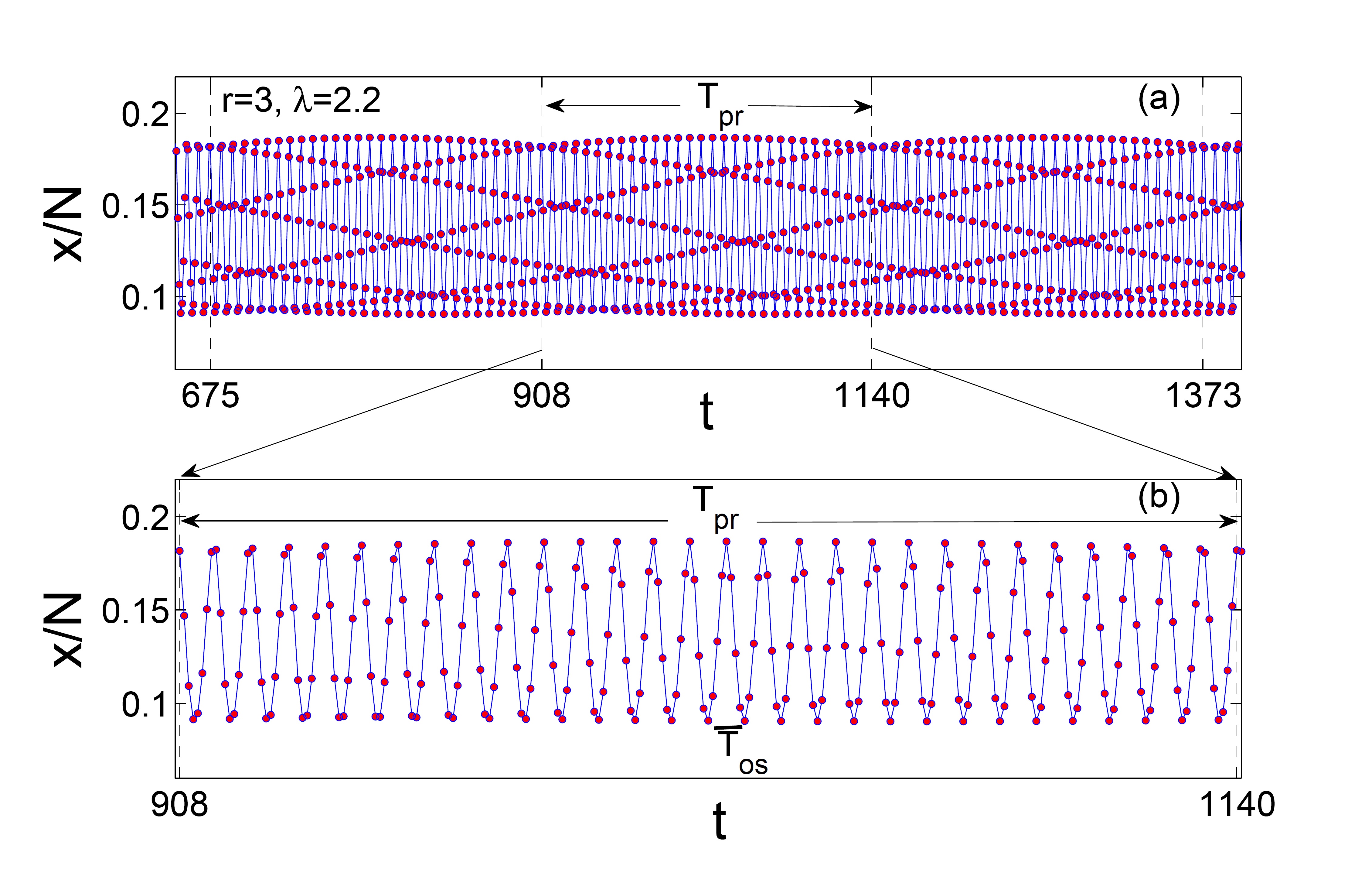}
    \caption{(Color online) Mean-field solution for oscillating phase in a system with $r=3$, $\lambda=2.2$ and $N=40000$. (a) Slow oscillations with large period of $T_{pr}=232$. (b) Fast oscillations with period $T_{os}=8$.}
    \label{Fig12}
\end{figure}

\begin{figure}[!htbp]
\centering
    \includegraphics[width=0.9\linewidth]{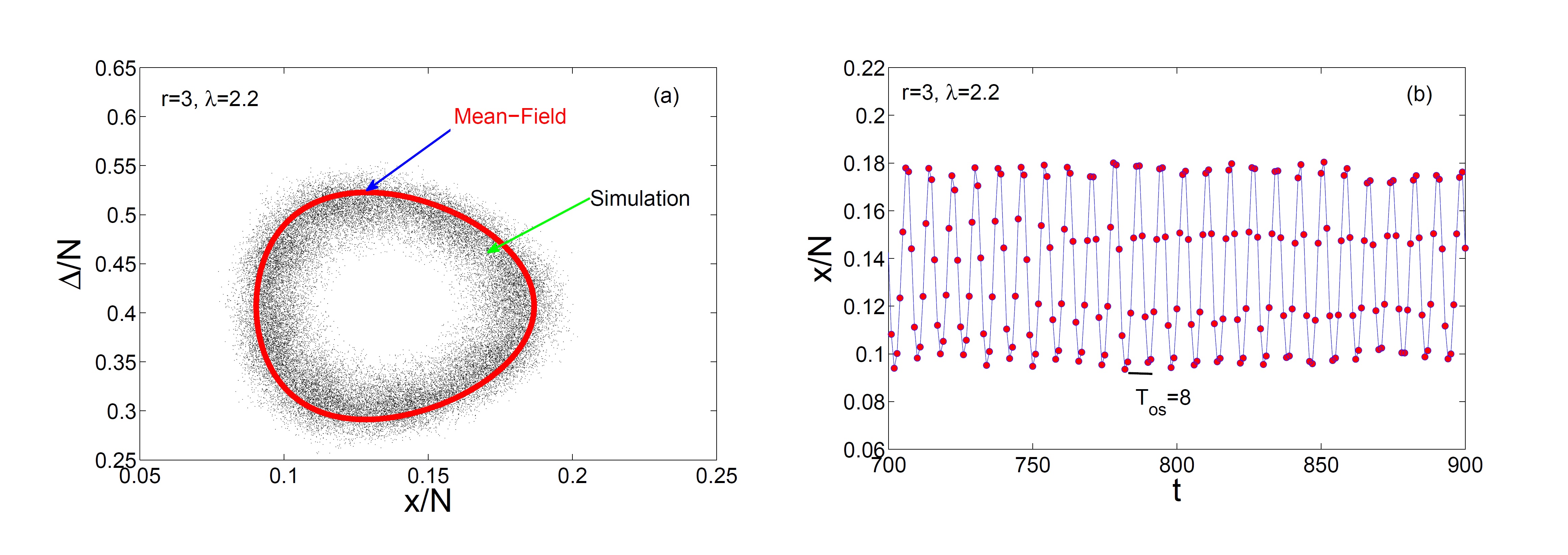}
    \caption{(Color online) Simulation results of the oscillating phase for a system with $r=3$, $\lambda=2.2$ and $N=40000$. (a) The accessible region of the phase space for the dynamics. Red line is the mean-field approximation and the dotted area is obtained from simulations. (b) Oscillations with period $T_{os}=8$ and fluctuating amplitude obtained from numerical simulations.}
    \label{Fig13}
\end{figure}

An example of oscillatory behavior is presented in Fig. \ref{Fig11}, for $\lambda=2.2$ and $r=3$ the dynamics periodically jumps between eight points that are rotating on an egg-shaped attractor in the phase space. This leads to fast oscillations with period $T_{os}=8$ that is the result of jumps between the eight points. This fast oscillation, due to rotation around the attractor, is embedded in another slow oscillation with a larger period of $T_{pr}=29\times T_{os}$, see Fig. \ref{Fig12}. Here, we must note that the period of fast oscillations $T_{os}$ just depends on $r$ and is equal to $T_{os}=2(1+r)$, but $T_{pr}$ depends on $r$ as well as $\lambda$. To better understand this oscillatory behavior we performed numerical simulations of the actual dynamics described by Eq. (1) In Fig. \ref{Fig13} we present our results obtained from numerical simulation of the system with $\lambda=2.2$ and $r=3$. In Fig. \ref{Fig13}(a) it is clearly seen that, because of fluctuations, the one dimensional egg-shaped mean-field attractor is extended to a two dimensional area in the phase space. Fig. \ref{Fig13}(b) shows that, because of fluctuations, no slow oscillation ($T_{pr}$) is observed and fast oscillations with period $T_{os}=8$ with fluctuating amplitude dominate the activity of the oscillating phase. Note that $T_{os}=8$ is in agreement with the mean-field approximation. Oscillatory behavior has previously been reported in epidemic models with refractory period \cite{Smith,Hethcote}, their approach is based on solving integrodifferential master equations with continuous variables. Above we present a more simple study of such oscillations using the activity dependent branching ratio and a discrete phase space. However, our focus is mainly on the parameter regime where we expect to observe critical behavior.

\begin{figure}[!htbp]
\centering
    \includegraphics[width=0.9\linewidth]{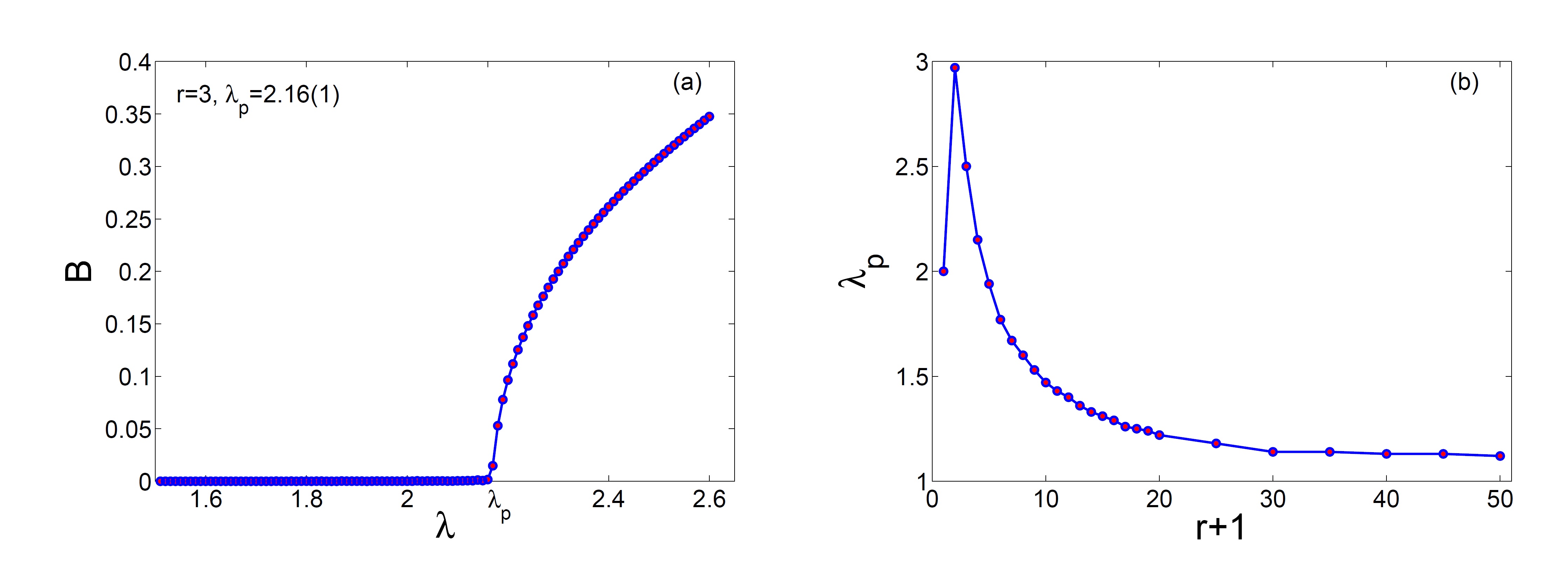}
    \caption{(a) Plot of the amplitude of oscillations as a function of $\lambda$. Zero amplitude is related to stable fixed points and non-zero amplitude is related to the oscillating phase. At $\lambda=\lambda_{p}$ the fixed point becomes unstable and a transition to the oscillating phase takes place. (b) Plot of $\lambda_{p}$ as a function of $r+1$ (number of refractory time steps). }
    \label{Fig14}
\end{figure}

Now, we focus on the transition point $\lambda_{p}$ in which the single point attractor becomes unstable and transition to the oscillating state takes place. In order to find the transition point we calculate the amplitude of oscillations ($B$) for each value of $\lambda$. In the case of one single attractor $B$ is equal to zero. In the case of oscillatory behavior $B$ becomes larger than zero.  By plotting $B$ versus $\lambda$ we can find $\lambda_{p}$, see Fig. \ref{Fig14}(a). Fig. \ref{Fig14}(b) shows $\lambda_{p}$ for different values of $r$, we see $\lambda_{p}>1$ even for very large refractory times. We conclude that there exist a range of $1\leq\lambda<\lambda_{p}$ where the system exhibits one attractor at which $b(x^{*},\Delta^{*})=1$ and we expect to observe critical behavior.


\newpage

\begin{thebibliography}*
\bibitem{BuK} Burridge R. $\&$ Knopoff, L. Model and theoretical seismicity. \textit{Bull. Seismol. Soc. Am.} \textbf{57}, 341-371 (1967).
\bibitem{OFC} Olami, Z., Feder, H. J. S. $\&$ Christensen, K. Self-organized criticality in a continuous, nonconservative cellular automaton modeling earthquakes. \textit{Phys. Rev. Lett.} \textbf{68}, 1244 (1992).
\bibitem{BP1} Beggs, J. M. $\&$ Plenz, D. Neuronal avalanches in neocortical circuits. \textit{J. Neurosci.} \textbf{23}, 11167-11177 (2003).
\bibitem{SYYRP} Shew, W. L., Yang, H., Yu, S., Roy, R. $\&$ Plenz. D., Information capacity and transmission are maximized in balanced cortical networks with neuronal avalanches. \textit{J. Neurosci.} \textbf{31}, 55-63 (2011).
\bibitem{SYPRP} Shew, W. L., Yang, H., Petermann, T., Roy, R. $\&$ Plenz. D. Neuronal avalanches imply maximum dynamic range in cortical networks at criticality \textit{J. Neurosci.} \textbf{29}, 15595-15600 (2009).
\bibitem{HB}Haldeman, C., $\&$ Beggs, J. M. Critical branching captures activity in living neural networks and maximizes the number of metastable states. \textit{Phys. rev. lett.} \textbf{94}, 058101 (2005).
\bibitem{Garcia}Williams-Garc{\'i}a, R. V., Moore, M., Beggs, J. M., $\&$ Ortiz, G. Quasicritical brain dynamics on a nonequilibrium widom line. \textit{Phy. Rev. E}, \textbf{90}, 062714  (2014).
\bibitem{BTW1} Bak, P., Tang, C., $\&$ Wiesenfeld, K. Self-organized criticality: An explanation of the $1/f$ noise. \textit{Phys. Rev. Lett.} \textbf{59}, 381 (1987).
\bibitem{BTW2} Bak, P., Tang, C., $\&$ Wiesenfeld, K. Self-organized criticality. \textit{Phys. Rev. A} \textbf{38}, 364 (1988).
\bibitem{B}  Bak, P. \textit{How nature works: the science of self-organized criticality} (Springer-Verlag New York, 1996).
\bibitem{P}  Pruessner, G. \textit{Self-organized criticality: theory, models and characterisation} (Cambridge University Press New York, 2012).
\bibitem{Grassberger} Son, S. W., Bizhani, G., Christensen, C., Grassberger, P. $\&$ Paczuski, M. Percolation theory on interdependent networks based on epidemic spreading. \textit{Europhys. Lett.} \textbf{97}, 16006 (2012).
\bibitem{Karrer} Karrer, B. $\&$ Newman, M. E. J. Competing epidemics on complex networks. \textit{Phys. Rev. E} \textbf{84}, 036106 (2011).
\bibitem{Mieghem}  Van Mieghem, P. Epidemic phase transition of the SIS type in networks. \textit{Erophys. Lett.} \textbf{97}, 48004 (2012).
\bibitem{Montakhab} Montakhab, A. $\&$ Manshour, P. Low prevalence, quasi-stationarity and power-law behavior in a model of contagion spreading. \textit{Erophys. Lett.} \textbf{99}, 58002 (2012).
\bibitem{Manshour} Montakhab, A. $\&$ Manshour, P. Contagion spreading on complex networks with local deterministic dynamics. \textit{Commun. Nonlinear Sci. Numer. Simul.} \textbf{19}, 2414-2422 (2014).
\bibitem{Plenz} Plenz, D. ed. \textit{Criticality in Neural Systems} (New York, NY: Wiley-VCH, 2014)
\bibitem{PT}  Plenz, D. $\&$ Thiagarajan, T. C. The organizing principles of neuronal avalanches: cell assemblies in the cortex?. \textit{Trends in Neurosci.} \textbf{30}, 101-110 (2007).
\bibitem{BP2}    Beggs, J. M. $\&$ Plenz, D. Neuronal avalanches are diverse and precise activity patterns that are stable for many hours in cortical slice cultures. \textit{J. Neurosci.} \textbf{24}, 5216-5229 (2004).
\bibitem{FIBSLD} Friedman, N. \textit{et. al.} Universal critical dynamics in high resolution neuronal avalanche data. \textit{Phys. Rev. Lett.} \textbf{108}, 208102 (2012).
\bibitem{PTLNCP} Petermann, T. \textit{et. al.} Spontaneous cortical activity in awake monkeys composed of neuronal avalanches.  \textit{Proc. Natl. Acad. Sci. USA} \textbf{106}, 15921-15926 (2009).
\bibitem{TBFC} Tagliazucchi, E., Balenzuela, P., Fraiman, D. $\&$ Chialvo, D. R. Criticality in large-scale brain fMRI dynamics unveiled by a
novel point process analysis. \textit{Front. Physiol.} \textbf{3}, 15 (2012).
\bibitem{SACHHSCBP} O. Shriki, \textit{et. al.} Neuronal avalanches in the resting MEG of the human brain. \textit{J. Neurosci.} \textbf{33}, 7079-7090 (2013).
\bibitem{HTBC}  Haimovici, A., Tagliazucchi, E., Balenzuela, P. $\&$ Chialvo, D. R. Brain organization into resting state networks emerges at criticality on a model of the human connectome. \textit{Phys. Rev. Lett.} \textbf{110}, 178101 (2013).
\bibitem{C}   Chialvo, D. Emergent complex neural dynamics. \textit{Nature Physics} \textbf{6}, 744-750 (2010).
\bibitem{BN}  Beggs, J. M. $\&$ Timme, N. Being critical of criticality in the brain. \textit{Front. in physiol.} \textbf{3}, 163 (2012).
\bibitem{DanteC} Chialvo, D. Critical brain networks.  \textit{Physica A} \textbf{340}, 756-765 (2004).
\bibitem{BR} Bornholdt, S. $\&$ R\"{o}hl, T. Self-organized critical neural networks \textit{Phys. Rev. E} \textbf{67}, 066118 (2003).
\bibitem{LHG} Levina, A., Herrmann, J. M. $\&$ Geisel. T. Dynamical synapses causing self-organized criticality in neural networks. \textit{Nature Physics} \textbf{3}, 857-860 (2007).
\bibitem{MMKN} Millman, D., Mihalas, S., Kirkwood, A. $\&$ Niebur. E. Self-organized criticality occurs in non-conservative neuronal networks during \lq up' states. \textit{Nature physics} \textbf{6}, 801-805 (2010).
\bibitem{APH} de Arcangelis, L., Perrone-Capano, C. $\&$ Herrmann, H. J. Self-Organized Criticality Model for Brain Plasticity. \textit{Phy. Rev. Lett.} \textbf{96}, 028107 (2006).
\bibitem{Beggs} Beggs, J. M. The criticality hypothesis: how local cortical networks might optimize information processing. \textit{Philos. Trans. R. Soc. A} \textbf{366}, 329-343 (2008).
\bibitem{AMIN} Moosavi, S. A. $\&$ Montakhab A. Mean-field behavior as a result of noisy local dynamics in self-organized criticality: Neuroscience implications. \textit{Phys. Rev. E} \textbf{89}, 052139 (2014).
\bibitem{AMIN1}  Moosavi, S. A. $\&$ Montakhab A. Structural versus dynamical origins of mean-field behavior in a self-organized critical model of neuronal avalanches. \textit{Phys. Rev. E} \textbf{92}, 052804 (2015).
\bibitem{AH} de Arcangelis, L. $\&$ Herrmann. H. J. Learning as a phenomenon occurring in a critical state. \textit{Proc. Natl. Acad. Sci. USA} \textbf{107}, 3977-3981 (2010).
\bibitem{KC} Kinouchi, O. $\&$ Copelli. M. Optimal dynamical range of excitable networks at criticality. \textit{Nature Physics} \textbf{2}, 384-351 (2006).
\bibitem{LSR1} Larremore, D. B., Shew, W. L. $\&$ Restrepo, J. G. Predicting criticality and dynamic range in complex networks: effects of topology. \textit{Phys. Rev. Lett.} \textbf{106}, 058101 (2006).
\bibitem{LSR2} Larremore, D. B., Shew, W. L., Ott, E. $\&$ Restrepo, J. G. Effects of network topology, transmission delays, and refractoriness on the response of coupled excitable systems to a stochastic stimulus. \textit{Chaos} \textbf{21}, 025117 (2011).
\bibitem{PTYJZZ}  Pei, S. \textit{et. al.} How to enhance the dynamic range of excitatory-inhibitory excitable networks. \textit{Phys. Rev. E} \textbf{86}, 021909 (2012).
%\bibitem{WXW} A. C. Wu, X. J. Xu, and Y. H. Wang, Phys. Rev. E \textbf{75}, 032901 (2007).
%\bibitem{GKC} L. L. Gollo, O. Kinouchi, PLoS Comput. Biol. \textbf{5}, e1000402 (2009).
\bibitem{Kandel} Kandel, E. R., Schwartz, J. H., Jessel, T. M., Siegelbaum, S. A. $\&$ Hudspeth, A. J. \textit{Principles of neural science, $5^{th}$ ed. (The McGraw-Hill Companies, 2013).}
\bibitem{LCOR} Larremore, D. B., Carpenter, M. Y., Ott, E. $\&$ Restrepo, J. G. Statistical properties of avalanches in networks. \textit{Phys. Rev. E} \textbf{85}, 066131 (2012).
\bibitem{LSOSR}  Larremore, D. B., Shew, W. L., Ott, E., Sorrentino, F. $\&$ Restrepo, J. G. Inhibition causes ceaseless dynamics in networks of excitable nodes. \textit{Phys. Rev. Lett.} \textbf{112}, 138103 (2014).
\bibitem{ROH} Restrepo, J. G., Ott, E. $\&$ Hunt, B. R. Approximating the largest eigenvalue of network adjacency matrices. \textit{Phys. Rev. E} \textbf{76}, 056119 (2007).
\bibitem{MSP}  Martin, E., Shreim, A. $\&$ Paczuski, A. Activity-dependent branching ratios in stocks, solar x-ray flux, and the Bak-Tang-Wiesenfeld sandpile model. \textit{Phys. Rev. E} \textbf{81}, 016109 (2010).
\bibitem{A}     Alstr{\o}m, P. Mean-field exponents for self-organized critical phenomena. \textit{Phys. Rev. A} \textbf{38}, 4905 (1988).

\bibitem{Smith} Smith, H. On periodic solutions of a delay integral equation modeling epidemics, \textit{J. Math. Biol.} \textbf{4}, 69-80 (1977).
\bibitem{Hethcote} Hethcote, H. W., Stech, H. W., $\&$ Driessche, P. V. D. Nonlinear oscillations in epidemic models. \textit{SIAM J. Appl. Math.} \textbf{40}, 1-9 (1981).
\bibitem{Pinho} Pinho, S. T. R., $\&$ Andrade, R. F. S., Power-law sensitivity to initial conditions for Abelian directed self-organized critical models. \textit{Physica A} \textbf{344}, 601-607 (2004).

\end{thebibliography}
\end{document}